# COVID-19 Susceptibility, Mortality and Length of Hospitalization based on Age-Sex Composition: Insights for Intervention and Stratification

Roel F. Ceballos
Mathematics and Statistics Department
University of Southeastern Philippines
Davao City, 8000 Philippines
*roel.ceballos@usep.edu.ph*

## Abstract

*The coronavirus disease (COVID-19) has spread worldwide with an unprecedented impact on society. In the Philippines, several interventions such as mobility restrictions for different age groups and vaccination prioritization programs have been implemented to reduce the risks of infections and mortality. This study aimed to identify age-sex composition with greater susceptibility, longer hospitalization and higher fatality. The COVID-19 cases from March 2020 to April 2021 provided by the Department of Health Davao Region in the Philippines were analyzed. A Chi-square test was used to determine the difference in proportions of COVID-19 cases among age-sex compositions. A correlation plot of $\chi^2$ test residual was employed to investigate the differences in susceptibility. Boxplots and Kruskal-Wallis tests were utilized to compare the length of hospitalizations. The study found a significant difference in the COVID-19 susceptibility among age-sex compositions (p < 0.01). Male children and female senior citizens were the most susceptible age-sex compositions. Furthermore, senior citizens had the longest hospital days wherein the median and IQR days were 19 (15-27) for men and 18 (16-29) for women. Male senior citizen was the subgroup with the highest case fatality (21.4%, p < 0.01). It is recommended that the number of cases among senior citizens be used as an input in the planning and allocation of medical resources at the provincial and regional levels. The local government unit executives in the region can also take advantage of the availability of age-sex composition data in stratifying localities, planning, allocating COVID-19-related resources and imposing mobility restrictions.*

*Keywords*: susceptibility, demographic subgroup, vaccination priority, sex and age

## 1. Introduction

The coronavirus disease (COVID-19) is highly transmittable caused by severe acute respiratory syndrome coronavirus 2 (SARS-CoV-2), which reportedly emerged in Wuhan, China (Shereen *et al*., 2020). On January 30, the World Health Organization announced a Public Health Emergency of International Concern for the COVID-19 outbreak (ul Qamar *et al.*, 2020). As of June 3, 2021, the virus has spread to almost 200 countries and territories including 26 cruise and naval ships (Mallapaty, 2020). In the Philippines, several interventions such as restrictions on mobility and vaccination prioritization programs according to age groups have been implemented to reduce the risk of infection and mortality. The Inter-Agency Task Force (IATF), an advisory body responsible for decisions on the management and control of COVID-19 in the Philippines, has rolled out priority groups for the COVID-19 vaccination program to reduce mortality, preserve health system capacity and protect the most-at-risk populations (Department of Health [DOH], 2020). Workers in frontline health services are the top priority for vaccination followed by senior citizens and persons with comorbidities.

On the other hand, community or city lockdowns and mobility restrictions have also been implemented by IATF from time to time, especially when there are spikes in the number of cases and deaths (IATF, 2020). As of 24 August 2021, the Philippines has accumulated 1.87 million cases with 32,264 deaths. The average number of positive cases per day is high, around 14,859.

Identifying age-sex composition with higher susceptibility and mortality will enable health experts and policymakers to take more effective surveillance and targeted interventions to minimize the pandemic's adverse effects. Earlier epidemiological studies showed that sex and age are two of the most important factors influencing the susceptibility and severity of COVID-19 (Chen *et al*., 2020; Dudley *et al*., 2020; Jin *et al*., 2020; Ceballos, 2021). Older age was associated with higher severity and mortality, and the number of men who died from COVID-19 was 2.4 times higher than that of women (Dudley *et al*., 2020). However, these studies have mainly concentrated on large economies. Little is known whether the reported age-sex dependent patterns exist in low-income/middle countries, explicitly considering data within community levels and local government units. In this regard, this study was conducted to determine the age-sex patterns of COVID-19 susceptibility, mortality and length of hospitalization in the Davao Region, Philippines to provide insights to local government units. Davao Region was chosen because it has the largest population (5,290,869 as of 2020) in Mindanao. As such, it is one of the fastest-growing regional economies in the Philippines (Philippine Statistics Authority [PSA], 2021).

## 2. Methodology

*2.1 Study Design and Data Variables*

A retrospective design was employed involving COVID-19 positive cases in Davao Region from March 2020 to April 2021. The data was obtained from the report of the Davao Center for Health Development of the DOH, Davao Region. The report is released to the public daily through their official Facebook page. Furthermore, to ensure the anonymity of each case, the dataset contained no personal information or patient identifier. It comprised a total of 24,295 cases containing the following information: age, sex, provincial residence, date of admission to either hospital or quarantine facility, and the status of whether recovered or died or still admitted at the time of data retrieval. Age was categorized into four, namely children (0-12 years old), teens (13 to 19 years old), adults (20 to 59 years old), and senior citizens (60 years old and above). Survival days were calculated based on the date of diagnosis using the reverse transcription-polymerase chain reaction test. Mortality status was categorized as dead or alive as of 20 May 2021 – the date of data retrieval. The age-sex composition mortality rates due to COVID-19 were computed using the population by age-sex composition as the denominator. The population data was obtained from the publication of the PSA, a government agency responsible for developing and conducting population censuses in the country (PSA, 2020).

*2.2 Statistical Analysis*

All categorical variables were presented as numbers (*n*) and percentages (%). A Chi-square test was used to determine the difference in proportions of COVID-19 cases among age-sex compositions. A correlation plot of $\chi^2$ test residual was used to investigate the differences in susceptibility. The male to female sex ratio was also utilized to compare the susceptibility and mortality of the different age groups by province in the region. Case fatality ratio was also computed in addition to the methods used to measure the mortality of COVID-19. Boxplot was used to display the hospitalization days and survival days of COVID-19 patients. It showed the minimum, maximum, median, and first and third quartiles of the datasets. Kruskal-Wallis test was employed to investigate the differences in the median hospitalization days and survival days of COVID-19 patients by age-sex compositions. For all tests, a two-tailed P-value of < 0.05 was considered statistically significant. Statistical analysis was carried out using the R Programming language (R Core Team, 2013).

## 3. Results and Discussion

A total of 24,295 COVID-19 cases (12,197 men and 12,098 women) were recorded in this region during the study period (Table 1). The majority of deaths were recorded among men (60%). For the age groups, most COVID-19 cases were recorded among the adult's group (73.9%), but most deaths were among senior citizens (55.5%). Furthermore, Davao del Sur had the highest recorded cases (68.1%) and deaths (75.7%) among the provinces in Davao Region.

Table 1. Summary of COVID-19 cases and deaths in Davao Region

| Characteristics | Cases $n$ (%) | Deaths $n$ (%) |
|---|---|---|
| Sex | | |
| Male | 12,197 (50.2) | 602 (60.0) |
| Female | 12,098 (49.8) | 402 (40.0) |
| Age, median (IQR) | | |
| Children (0-12 years old) | 1,602 (6.6) | 26 (2.6) |
| Teens (13-19 years old) | 1,692 (7.0) | 9 (0.9) |
| Adults (20-60 years old) | 17,952 (73.9) | 412 (41.0) |
| Senior citizens ($\geq$ 60 years old) | 3,049 (12.5) | 557 (55.5) |
| Provinces in Davao Region | | |
| Davao del Sur | 16,554 (68.1) | 760 (75.7) |
| Davao del Norte | 4,381 (18.0) | 149 (14.8) |
| Davao de Oro | 1,658 (6.8) | 50 (5.0) |
| Davao Oriental | 1,466 (6.0) | 36 (3.6) |
| Davao Occidental | 236 (1.0) | 9 (0.9) |

*3.1 Susceptibility by Sex and Age composition*

Table 2 shows the number and percentage of cases for the different age-sex compositions from March 2020 to April 2021 in the region. The adult group had the highest number of cases among men (74.0%) and women (73.8%) followed by senior citizen women (13.1%) and men (12.0%).

Table 2. COVID-19 cases by sex and age group from March 2020 to April 2021

| Sex group | Age group | | | |
|---|---|---|---|---|
| | Children | Teens | Adults | Senior citizens |
| Men | | | | |
| Cases, $n$ (%) | 870 (7.1) | 838 (6.9) | 9,029 (74.0) | 1,460 (12.0) |
| Women | | | | |
| Cases, $n$ (%) | 732 (6.1) | 854 (7.1) | 8,923 (73.8) | 1589 (13.1) |

The proportions of COVID-19 cases among age-sex combinations were significantly different ($\chi^2$ test, $p < .01$) (Table 2) implying that some subgroups were more susceptible than others.

The proportions of COVID-19 cases in women were higher than 50% among teens and senior citizens, while male cases were higher than 50% among children and adults. The difference in proportions of COVID-19 cases among age-sex combinations was significant ($\chi^2$ test, $p < 0.01$) (Figure 1a), which suggested that some subgroups were more susceptible than others. To further investigate these differences in susceptibility, the residuals of the $\chi^2$ test were presented using a correlation plot (Figure 1b), which reveals that male children had the highest susceptibility among COVID-19 cases in the Davao Region (indicated by the large dark blue circle). Female senior citizens were second-most susceptible to COVID-19 infection than other age and sex compositions. Furthermore, female children were the least susceptible subgroup (indicated by the large dark red-brown circle).

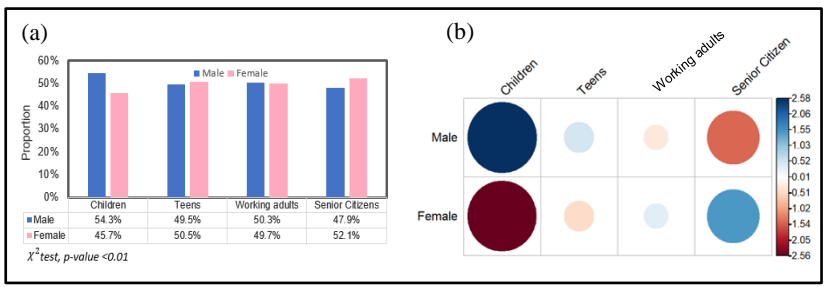

Figure 1. COVID-19 cases by sex and age groups

It was interesting to know if the same pattern existed at provincial levels. For this purpose, the male-to-female sex ratio of COVID-19 cases for the different age groups by province in the Davao Region is shown in Table 3. Davao Region is composed of five provinces, namely Davao del Sur, Davao del Norte, Davao de Oro, Davao Oriental and Davao Occidental. It was observed that across all provinces in the region, male children had greater susceptibility than female children (SR: 1.03 to 1.58). This result was consistent with the correlation plot in Figure 1b. Furthermore, lower susceptibility of male senior citizens was observed in Davao de Oro and Davao Oriental. In contrast, a close male-to-female ratio was noticed in Davao del Sur and Davao Del Norte.

Table 3. Sex ratio of COVID-19 cases per 100,000 population

| Province | Age group | | | |
|---|---|---|---|---|
| | Children | Teens | Adults | Senior Citizens |
| Davao del Sur | 1.12 | 1.09 | 1.00 | 1.07 |
| Davao del Norte | 1.03 | 0.78 | 0.80 | 1.04 |
| Davao de Oro | 1.58 | 0.79 | 0.81 | 0.92 |
| Davao Oriental | 1.03 | 0.77 | 0.77 | 0.80 |
| Davao Occidental | 1.21 | 1.11 | 1.29 | 1.22 |

The results revealed notable differences in susceptibility among different age groups. Male children were highly susceptible to COVID-19 infections compared with other age-sex compositions. This pattern existed across provincial levels. According to Bwire (2020), males are more vulnerable to infections due to sex-based immunological differences. Furthermore, Simon *et al.* (2015) stressed that children below five years old have low immunity and are at higher risk of infections. Despite being very susceptible to COVID-19 and other viral infections, children are very much adapted to respond and have a lower risk of mortality than other age groups (Nogrady, 2020).

*3.2 Length of Hospitalization*

In Tables 4 and Figure 2, the length of hospitalization for the recovered group or those patients who were considered recovered at the time of data retrieval is presented. The length of hospitalization for females was statistically different for all age groups ($p < 0.0001$) (Figure 2b). Furthermore, senior citizens with a median and IQR hospital days of 19 (15-27) had the longest hospital days followed by children and adults with a median and IQR hospital days of 13 (11-15) and 13 (10-17), respectively. The result was almost the same for the male group, except that the length of hospitalization for children and teens was not statistically different (Figure 2a).

Table 4. Length of hospitalization for men and women

| Sex group | Age group | | | | P-value |
|---|---|---|---|---|---|
| | Children | Teens | Adults | Senior citizens | |
| Men | | | | | |
| Median | 13 | 11 | 13 | 19 | < 0.001 |
| Interquartile range | 11-15 | 9-16 | 11-18 | 15-27 | |
| Women | | | | | |
| Median | 13 | 11 | 13 | 18 | < 0.001 |
| Interquartile range | 11-15 | 8-15 | 10-17 | 16-29 | |

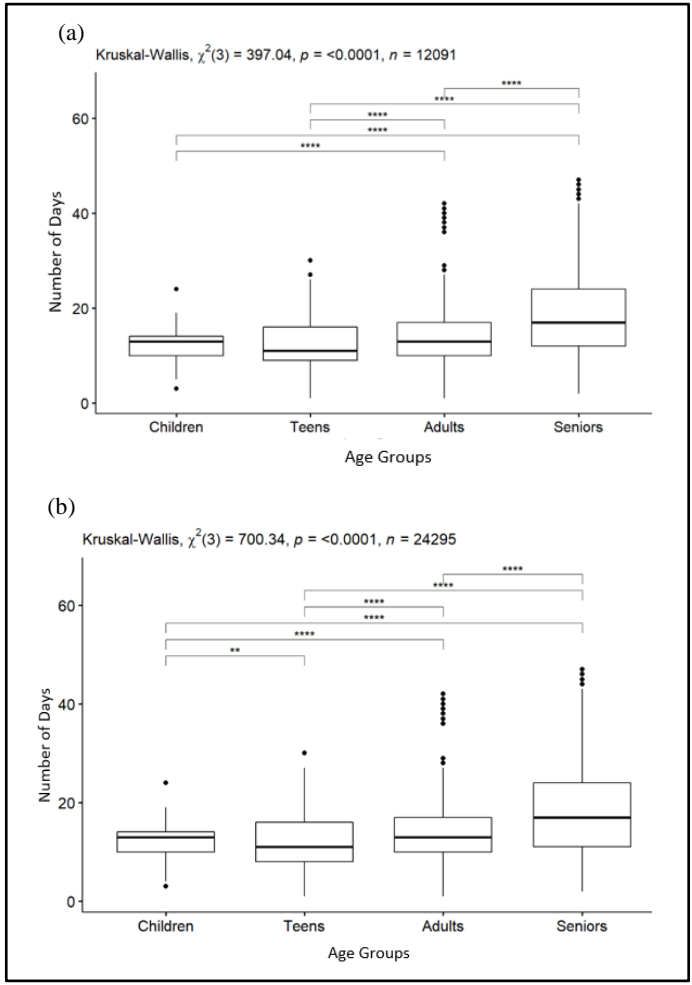

Figure 2. Hospitalization days for recovered COVID-19 cases: men (a) and women (b) groups

The results exhibited significant differences in the length of hospital stay among different age groups. Interestingly, studies conducted during the early onset of COVID-19 found little to no statistical difference in the length of hospital stay among age groups (Qiu *et al.*, 2020; Shi *et al.*, 2020; Wei *et al.*, 2020; Xia *et al.*, 2020). However, recent studies showed that age is a significant factor in longer hospitalization and that similar to the findings of this study, senior citizens had the longest hospital stay (Chiam *et al.*, 2021; Guo *et al.*, 2021; Keegan and Lysons, 2021). According to these studies, aging COVID-19 patients have greater need for critical care due to COVID-19-related comorbidities. Their probability of staying in the hospital was high compared with other age groups.

*3.3 Case Fatality Rate and Mortality*

Table 5 discloses that most reported deaths were among senior citizens – 51.8% and 60.9% for men and women, respectively. The result was followed by adults (men: 45% and women: 36%). Males' crude case-fatality ratio was significantly higher than females in adults and senior citizens. Furthermore, mortality per 100,000 population directly caused by COVID-19 among males was significantly higher than among females for all age groups except for teens (proportion test, $p < 0.01$) (Table 5).

Table 5. COVID-19 fatality and mortality rates between sex and age groups of

| Sex group | Age group | | | |
|---|---|---|---|---|
| | Children | Teens | Adults | Senior citizens |
| Deaths, *n* (%) | | | | |
| Men | 17 (2.8) | 5 (0.8) | 268 (44.5) | 312 (51.8) |
| Women | 9 (2.2) | 4 (1.0) | 144 (35.8) | 245 (60.9) |
| Case fatality ratio (%) | | | | |
| Men | 2.0 | 0.6 | 3.0 | 21.4 |
| Women | 1.23 | 0.47 | 1.61 | 15.42 |
| Proportions test P-value | 0.052 | 0.629 | < 0.01 | < 0.01 |
| Mortality/100,000 population | | | | |
| Men | 1.23 | 0.65 | 9.91 | 71.15 |
| Women | 0.65 | 0.52 | 5.33 | 55.87 |
| Proportions test P-value | < 0.01 | 0.518 | < 0.01 | < 0.01 |

These findings were investigated by considering the male-to-female sex mortality ratio in each age group by province in the region. Table 6 shows that the sex mortality ratio of COVID-19 cases for each province was higher for men among senior citizens ranging from 1.01 to 1.54 in different provinces, which demonstrated that male senior citizens had higher mortality than other age-sex groups. Furthermore, the sex mortality ratio for male children and adults in Davao del Sur was higher than in other sex and age groups.

As observed in the analysis, male and female senior citizens obtained the highest risk of mortality and fatality. The result agrees with existing studies underscoring that mortality and fatality are highly associated with older age among COVID-19 patients. (Ho *et al*., 2020; Kang and Jung, 2020; Yanez *et al*., 2020).

Table 6. Sex ratio of mortality rates

| Province | Age group | | | |
|---|---|---|---|---|
| | Children | Teens | Adults | Senior Citizens |
| Davao del Sur | 1.87 | 1.23 | 1.92 | 1.54 |
| Davao del Norte | 0.95 | 0.95 | 1.41 | 1.30 |
| Davao de Oro | 1.00 | 0.93 | 1.77 | 1.05 |
| Davao Oriental | 1.05 | 0.93 | 0.90 | 1.01 |
| Davao Occidental | 0.94 | 0.96 | 0.93 | 1.10 |

## 4. Conclusion and Recommendation

The results revealed significant differences in the COVID-19 susceptibility, mortality and length of hospital days among the age-sex compositions in the Davao Region. Male children and female senior citizens were the most susceptible age-sex compositions. Hospital stay was longer for senior citizens regardless of sex, who also had the highest case of fatality and mortality rates.

The differences imply that age-sex compositions can be used to guide local government executives in the region in making immediate decisions from time to time in risk-stratifying localities, allocating and managing COVID-19 resources, and controlling mobility. Although vaccination already commenced across all populations (excluding children aged below five years old as of this writing), there is a need to boost campaign efforts to increase the vaccination rate, thereby reducing the risk of infection and death, especially among senior citizens.